\documentclass[prl, nofootinbib, superscriptaddress, preprintnumbers, twocolumn, showkeys]{revtex4}

\usepackage{amsmath}	
\usepackage{amsthm}	
\usepackage{microtype}
\usepackage{amssymb}	
\usepackage{datetime}
\usepackage{graphicx}
\usepackage{color}
\usepackage{verbatim}
\usepackage{bm}

\usepackage{hyperref}

\definecolor{grey}{rgb}{0.4,0.4,0.4}
\definecolor{dullmagenta}{rgb}{0.4,0,0.4}
\definecolor{darkblue}{rgb}{0,0,0.4}
\definecolor{midblue}{rgb}{0,0,0.5}
\definecolor{midred}{rgb}{0.5,0,0}
\definecolor{orange}{rgb}{1,0.5,0}
\definecolor{lightbrown}{rgb}{0.75,0.5,0.25}
\definecolor{tan}{cmyk}{0.14,0.42,0.56,0}
\definecolor{djunglegreen}{cmyk}{0.99,0,0.52,0}
\definecolor{lightgreen}{rgb}{0,1,0}
\definecolor{olivegreen}{cmyk}{0.64,0,0.95,0.40}
\definecolor{midgreen}{rgb}{0.0,0.675,0.0}
\definecolor{darkgreen}{rgb}{0,0.5,0}

\newcommand{\vs}{\vspace}

\renewcommand{\.}{\hspace{0.5mm}}

\newcommand{\ra}{\ensuremath{\rightarrow}}

\newcommand{\Wrm}{\ensuremath{\mathrm{W}}}

\newcommand{\crm}{\ensuremath{\mathrm{c}}}

\newcommand{\srm}{\ensuremath{\mathrm{s}}}

\newcommand{\Ocal}{\ensuremath{\mathcal{O}}}
\newcommand{\Pcal}{\ensuremath{\mathcal{P}}}

\renewcommand{\d}{\ensuremath{\mathrm{d}}}

\newcommand{\eg}{e.g.}
\newcommand{\ie}{ie.}

\newcommand{\cf}{cf.}

\let\baraccent=\= 
\renewcommand{\=}[1]{\stackrel{#1}{=}} 

\theoremstyle{definition}

\theoremstyle{remark}

\settimeformat{ampmtime}

\usepackage{float}

\newcommand{\para}[1]{\par\vspace{2mm}\noindent{\emph{{#1}}}\,---\,}
\makeatletter
\DeclareRobustCommand{\rcite}[1]{%
  \rcite@aux#1,\@nil{#1}%
}
\def\rcite@aux#1,#2\@nil#3{%
  \if\relax#2\relax
    Ref.~\cite{#3}%
  \else
    Refs.~\cite{#3}%
  \fi
}
\makeatother

\begin{document}

\title{Uncertainties in primordial black-hole constraints on the primordial power spectrum}

\author{Yashar Akrami}
\email{akrami@lorentz.leidenuniv.nl}
\affiliation{Lorentz Institute for Theoretical Physics, Leiden University, P.O. Box 9506, 2300 RA Leiden, The Netherlands}

\author{Florian Kuhnel}
\email{florian.kuhnel@fysik.su.se}
\affiliation{The Oskar Klein Centre for Cosmoparticle Physics, Department of Physics, Stockholm University, AlbaNova, SE-106 91 Stockholm, Sweden}

\author{Marit Sandstad}
\email{marit.sandstad@astro.uio.no}
\affiliation{Nordita, KTH Royal Institute of Technology and Stockholm University, Roslagstullsbacken 23, SE-106 91 Stockholm, Sweden}

\begin{abstract} 
The existence (and abundance) of primordial black holes (PBHs) is governed by the power spectrum of primordial perturbations generated during inflation. So far no PBHs have been observed, and instead, increasingly stringent bounds on their existence at different scales have been obtained. Up until recently, this has been exploited in attempts to constrain parts of the inflationary power spectrum that are unconstrained by cosmological observations. We first point out that the simple translation of the PBH non-observation bounds into constraints on the primordial power spectrum is inaccurate as it fails to include realistic aspects of PBH formation and evolution. We then demonstrate, by studying two examples of uncertainties from the effects of critical and non-spherical collapse, that even though they may seem small, they have important implications for the usefulness of the constraints. In particular, we point out that the uncertainty induced by non-spherical collapse may be much larger than the difference between particular bounds from PBH non-observations and the general maximum cap stemming from the condition $\Omega \leq 1$ on the dark-matter density in the form of PBHs. We therefore make the cautious suggestion of applying only the overall maximum dark-matter constraint to models of early Universe, as this requirement seems to currently provide a more reliable constraint, which better reflects our current lack of detailed knowledge of PBH formation. These, and other effects, such as merging, clustering and accretion, may also loosen constraints from non-observations of other primordial compact objects such as ultra-compact minihalos of dark matter.
\end{abstract}

\keywords{primordial black holes, primordial power spectrum, dark matter}

\preprint{NORDITA-2016-128}

\maketitle

\para{Introduction}Cosmological observations, in particular those of the cosmic microwave background (CMB) anisotropies~\cite{Hinshaw:2012aka, Adam:2015rua}, place tight constraints on the properties of the primordial density (or curvature) fluctuations on large scales through the accurate measurements of the primordial power spectrum~\cite{Hinshaw:2012aka, Ade:2015lrj}. These measurements, however, probe only a relatively small range of scales, \ie, wave numbers between $k \sim10^{-3}~\text{Mpc}^{-1}$ and $k \sim 1~\text{Mpc}^{-1}$. Even though the measured power spectrum on these cosmological scales provides strong evidence in support of an inflationary phase~\cite{Starobinsky:1980te, Guth:1980zm, Linde:1981mu, Albrecht:1982wi} in the early Universe, and constrains various inflationary models and their parameters~\cite{Ade:2015lrj}, it only probes a small region of the inflaton potential. It has therefore been of importance to try to extend the constraints on the curvature power spectrum to a wider range of scales using other cosmological and astrophysical measurements. Power-spectrum constraints have been extended to $k \sim 10^{4}~\text{Mpc}^{-1}$ through measurements of the CMB spectral distortions~\cite{Chluba:2012we, Chluba:2013pya}, and to $k \sim 10^{4}~\text{Mpc}^{-1}$ using constraints on entropy production between Big Bang nucleosynthesis and today~\cite{Jeong:2014gna}, although these are currently only (fairly weak) upper bounds on the amplitude of the spectrum.

One important set of such additional constraints on small scales has been provided by non-observations of primordial black holes (PBHs)~\cite{Carr:1974nx, Carr:1975qj} that are expected to have formed in the early Universe when very large density perturbations collapsed. In most scenarios, these overdensities are of inflationary origin \cite{Hodges:1990bf, Carr:1993aq, Ivanov:1994pa}.\footnote{Many more possibilities for PBH formation exist, and we refer the interested reader to corresponding reviews (\eg~Ref.~\cite{Carr:2016drx}).} As soon as the overdense regions come back into causal contact after inflation, they collapse if they exceed a medium-specific threshold. PBHs form mainly in the radiation-dominated epoch, hence a radiation medium is considered. Because of the connection between the formation of PBHs and the amplitude of primordial fluctuations, observations or non-observations of PBHs could potentially further constrain inflationary dynamics on scales far smaller than the cosmological ones. PBHs have not been observed yet, and constraints from their non-observations have been used~\cite{Carr:1993aq, Carr:1994ar,Green:1997sz, Peiris:2008be, Josan:2009qn, Garcia-Bellido:2016dkw} to place (currently weak) upper limits on the amplitude of the power spectrum over a very wide range of scales (from $k\sim 10^{-2}~\text{Mpc}^{-1}$ to $k\sim 10^{23}~\text{Mpc}^{-1}$~\cite{Carr:1993aq}). Similar arguments are behind another type of constraints on the power spectrum at small scales, namely those from the non-observations of ultra-compact minihalos of dark matter (UCMHs)~\cite{Bringmann:2011ut, Aslanyan:2015hmi} that are expected to have formed shortly after matter-radiation equality when a perturbation with a very large amplitude but not large enough to collapse to a black hole enters the horizon~\cite{Berezinsky:2003vn, Ricotti:2009bs, Scott:2009tu}.

By using well-known effects, in this {\it Letter} we question the current use of constraints on the existence and abundance of PBHs (and similarly of UCMHs) to constrain the primordial power spectrum. We first demonstrate, via a simple example, that taking into account the known realistic aspects of the PBH (and UCMH) formation and evolution, such as the well-studied critical collapse, makes the simple translation of the constraints on their existence and abundance into constraints on the power spectrum inaccurate. This, in turn, implies that the local details of the current non-observation constraints are not reliable, as the critical-collapse effect washes them out. We then demonstrate that the effect of non-spherical collapse, as is already known, can induce uncertainties that are orders of magnitudes larger than the differences between direct constraints from non-observations of PBHs and the overall limit from the maximum total dark matter. This renders the former constraints practically useless. We therefore suggest the early-Universe model-builders to apply only this overall maximum dark-matter constraint to their models, as it currently provides the most reliable constraint on the primordial power spectrum from PBHs.

\para{Primordial black holes and power-spectrum constraints}The na{\"i}ve first estimate of PBH formation postulates that the holes formed this way would have a mass $M$ of the order of the mass $M_{\rm H}$ of a black hole of horizon size equal to the Universe's horizon at their time of formation. If the underlying primordial power spectrum is Gaussian, the simplest assumptions dictate that the fraction $\beta$ of the collapsed patches in the Universe at their time of formation is given by~\cite{Niemeyer:1997mt}
\begin{equation}
	\beta
		\approx
								\mathrm{Erfc}\!
								\left(
									\frac{\delta_{c}}{\sqrt{2\,\sigma\.}}
								\right)
								.
								\label{eq:betaToP}
\end{equation}
Here, $\delta_{c}$ is the critical overdensity, which, in radiation domination, is found to be approximately equal to $0.45$ (\cf~Ref.~\cite{Musco:2004ak}). Note that this value is essentially independent of the mass of the collapsing space-time region. In Eq.~\eqref{eq:betaToP}, $\sigma$ denotes the root-mean square of the primordial density power spectrum $P_{\delta}$. $\mathrm{Erfc}$ is the complementary error function $\mathrm{Erfc} \equiv 1 - \mathrm{Erf}$, with $\mathrm{Erf}$ being the standard error function. For values of $\beta$ between $0$ and $2$, this function is invertible, and hence it might appear reasonable that a constraint on the density of PBHs of a certain mass could be translated into a constraint on the primordial power spectrum (as has been, for instance, performed in Refs.~\cite{Green:1997sz, Peiris:2008be, Josan:2009qn, Garcia-Bellido:2016dkw}).\footnote{In the case of a non-Gaussian power spectrum, the functional form for $\beta$ is modified, but the function is still invertible over the same domain.} This procedure would open up a possible way to constrain the primordial power spectrum at much smaller scales than those accessible with CMB observations \cite{Hinshaw:2012aka, Ade:2015lrj}.

However, this na{\"i}ve procedure has been shown to be too simplistic and may lead to tremendous errors (\cf~Refs.~\cite{Kuhnel:2015vtw, Kuhnel:2016exn}). There are several important effects which need to be accounted for in order for the transformation between the primordial power spectrum and the PBH mass distribution to yield reliable answers.

\para{Sources of uncertainty}The first and best studied of these effects is perhaps that of the critical collapse. As has been argued theoretically \cite{Choptuik:1992jv, Koike:1995jm}, and was later also found in numerical investigations \cite{Niemeyer:1997mt, Musco:2004ak, Musco:2008hv, Musco:2012au} of collapse in full general relativity, primordial black holes actually form through the so-called {\it critical collapse}. By this, it is meant that the holes are not produced mono-chromatically, with their mass $M$ just being equal (or proportional) to the horizon mass, but form subject to the so-called {\it critical scaling}
\begin{align}
	M
		&=
								k\.M_{\rm H}
								\left(
									\delta
									-
									\delta_{c}
								\right)^{\gamma}
								\, .
								\label{eq:critical-scaling}
\end{align}
Here, $k$ is a real, positive constant, and the quantity $\delta$ denotes the overdensity. In radiation domination and for spherical density profiles{\,---\,}which will be assumed from now on{\,---\,}one finds $\gamma = 0.36$ and $k = 3.3$ (\cf~Ref.~\cite{Musco:2004ak}). Eq.~\eqref{eq:critical-scaling} describes the generation of a PBH mass distribution at each instance of time at which the overdensities reenter the horizon. Generically, also the initial spectrum of overdensities will be extended, leading to PBH formation in a range of different horizon masses, which will then be convoluted with the critical-collapse effect. Hence, {\it one looses the one-to-one correspondence between PBH mass and formation time/scale.}

If the shapes of the overdensities are non-spherical, which is to be expected in a realistic distribution, this may also lead to large effects, as has been pointed out recently in Ref.~\cite{Kuhnel:2016exn}. This non-sphericity effect is likely to strongly reduce the overall production of PBHs, yielding significantly weaker constraints on the primordial power spectrum from PBH non-observations. Although corresponding detailed numerical studies are still lacking, utilizing the estimate for ellipsoidal collapse of Ref.~\cite{Kuhnel:2016exn}, we are essentially led to changes in the threshold of the density contrast $\delta_{c}$, which increases as
\begin{align}
	\delta_{c}
		&\ra
								\delta_{ec}
		\equiv
								\delta_{c}\Bigg[
									1
									+
									\kappa\!
									\left(
										\frac{ \sigma^{2} }{ \delta_{\crm}^{2} }
									\right)^{\!\!{\nu}}
								\Bigg]
								\; .
								\label{eq:deltac->deltaec}
\end{align}
This is exactly the functional form as found in the study of ellipsoidal galactic halo formation \cite{Sheth:1999su}. The parameters $\kappa$ and $\nu$ are unknown, but suggestions can be the theoretical na{\"i}ve estimates $\kappa = 9 / \sqrt{10\pi}$ and $\nu = 1 / 2$, or halo-like estimates $\kappa = 0.47$ and $\nu = 0.67$. It is easy to show, by demanding the same final value of $\beta$ (\cf~Eq.~\eqref{eq:betaToP}), and given a model with a certain threshold $\delta_{c}$ and a variance $\sigma_{1}$, that changing to a new threshold $\delta_{ec}$ (\cf~Eq.~\eqref{eq:deltac->deltaec}) demands the replacement
\begin{align}
	\sigma_{1}
		&\ra
	\sigma_{2}
		\approx
								\frac{ \delta_{c} }
								{ \Wrm\Big( \exp\big[ - \delta_{c} / (2\.\sigma_{1}^{2} ) \big] 
								\delta_{ec}^{2} / \sigma_{1}^{2} \Big) }
								\; .
								\label{eq:Non-Spher}
\end{align}
Here, $\Wrm$ is the Lambert $\Wrm$-function, which is defined as the inverse function of $f( \Wrm ) = \Wrm \exp{( \Wrm )}$.

Another source of potential uncertainty is provided by primordial non-Gaussianities. These might be particularly relevant as PBH formation requires density contrasts of large, \ie~$\Ocal( 1 )$, magnitude deep inside the tail of the respective distribution where deviations from the Gaussian spectrum may have a particularly large effect. Like non-sphericities, these more or less shift the PBH abundance, albeit this effect can be in both directions, depending on the sign of the non-Gaussianity \cite{Byrnes:2012yx, Young:2014oea, Young:2015kda, Carr:2016drx}. However, if a significant amount of dark matter exists in the form of PBHs formed from a non-Gaussian spectrum, this will also add isocurvature effects to the CMB. As these are strongly constrained, PBHs from such an origin should be ruled out {\it a priori} as a large contributor to dark matter; for details see Refs.~\cite{Young:2014oea, Young:2015kda, Tada:2015noa}. Hence, at the current constraint level, non-gaussian effects of models will either have a relatively moderate effect, or the isocurvature effects they have will rule them out immediately. Hence, we do not consider such effects directly here.

Besides the mentioned effects, PBHs could undergo accretion \cite{1981MNRAS.194..639C} or merger events \cite{Meszaros:1975ef, Carr:1975qj, Meszaros:1980bf} that would also modify the spectrum observed today, and would thus also contribute to the mixing of PBH masses originating from different parts of the primordial overdensities.

Here, we argue that these realistic effects together, and maybe in particular the less well-studied effects, make it premature to use the non-observations of PBHs to reliably constrain the inflationary power spectrum, or any other PBH formation sources, at the level implied by the exclusion plots that are widely used at present. Although this may be known to the majority of PBH experts, such constraints may also be used by those working on inflation or other aspects of the early Universe. The detailed use of the plots reflecting details in observational constraints may imply to the outsider a level of certainty in the constraints, which we argue is highly overstated.

\para{Two examples: critical collapse and non-sphericities}Let us demonstrate this first by a concrete example where the inclusion of the critical collapse to a seemingly excluded power spectrum will render a PBH mass distribution allowed by the same data used to exclude it. We do this by incorporating the critical-collapse effect in our analysis. As an underlying set-up, which generates the initial spectrum of overdensities, we use the running-mass inflationary model \cite{Stewart:1996ey, Stewart:1997wg} as in Ref.~\cite{Carr:2016drx}. PBH formation in this model has been intensively studied in the literature (\cf~Refs.~\cite{Drees:2011yz, Drees:2011hb, Drees:2012sz, Leach:2000ea, Kuhnel:2015vtw, Carr:2016drx} and references therein). Perhaps the simplest realization of this model is via the inflationary potential
\begin{align}
	\label{pot1}
	V( \phi )
		&=
								V_{0}
								+
								\frac{ 1 }{ 2 }\.m_{\phi}^{2}(\phi)\.\phi^{2}
								\, ,
\end{align}
where $\phi$ is the inflaton and $V_{0}$ is a constant. There exists a plethora of embeddings of this model in various frameworks, such as hybrid inflation \cite{Linde:1993cn}, which lead to different functions $m_{\phi}( \phi )$. These yield distinct expressions for the primordial density power spectra whose variance can be recast into the general form \cite{Drees:2011hb}
\begin{align}
	\big[ \sigma( k ) \big]^{2}
		&\simeq
								\frac{ 8 }{ 81 }\.\Pcal( k_{\star} )
								\bigg(
									\frac{ k }{ k_{\star} }
								\bigg)^{\!\! n( k ) - 1}
								\Gamma\!
								\left(
									\frac{ n_{\srm}( k ) + 3 }{ 2 }
								\right)
								\label{eq:sigma-running-mass}
								,
\end{align}
where the spectral indices $n( k )$ and $n_{\srm}( k )$ are given by
\begin{subequations}
\begin{align}
	n( k )
		&=
								n_{\srm}( k_{\star} )
								-
								\frac{ 1 }{ 2! }\.\lambda_{1} \ln\!
								\bigg(
									\frac{ k }{ k_{\star} }
								\bigg)
								+
								\frac{ 1 }{ 3! }\.\lambda_{2} \ln^{2}\!
								\bigg(
									\frac{ k }{ k_{\star} }
								\bigg)
								\notag
								\\[0.5mm]
		&\phantom{=\;}
								-
								\frac{ 1 }{ 4! }\.\lambda_{3} \ln^{3}\!
								\bigg(
									\frac{ k }{ k_{\star} }
								\bigg)
								+
								\ldots
								\; ,
								\label{eq:n-running-mass}
								\displaybreak[1]
								\\[1mm]
	n_{\srm}( k )
		&=
								n_{\srm}( k_{\star} )
								-
								\lambda_{1}\.\ln\!
								\bigg(
									\frac{ k }{ k_{\star} }
								\bigg)
								+
								\frac{1}{2}\.\lambda_{2}\.\ln^{2}\!
								\bigg(
									\frac{ k }{ k_{\star} }
								\bigg)
								\notag
								\displaybreak[1]
								\\[0.5mm]
		&\phantom{=\;}
								-
								\frac{1}{6}\.\lambda_{3}\.\ln^{3}\!
								\bigg(
									\frac{ k }{ k_{\star} }
								\bigg)
								+
								\ldots
								\; ,
								\label{eq:n-running-mass-2}
\end{align}
\end{subequations}
with real parameters $\lambda_{i}$, $i = 1, 2, 3$. The spectral index and amplitude of the primordial power spectrum at the pivot scale $k_{\star} = 0.002\.{\rm Mpc}^{-1}$ have been measured \cite{Komatsu:2010fb, Planck:2013jfk, Ade:2015lrj} to be $n_{\rm s}( k_{\star} ) \approx 0.96 < 1$ and $\Pcal( k_{\star} ) = \Ocal( 10^{-9} )$.

\begin{figure}
	\vs{-3mm}
	\centering
	\includegraphics[scale=1,angle=0]{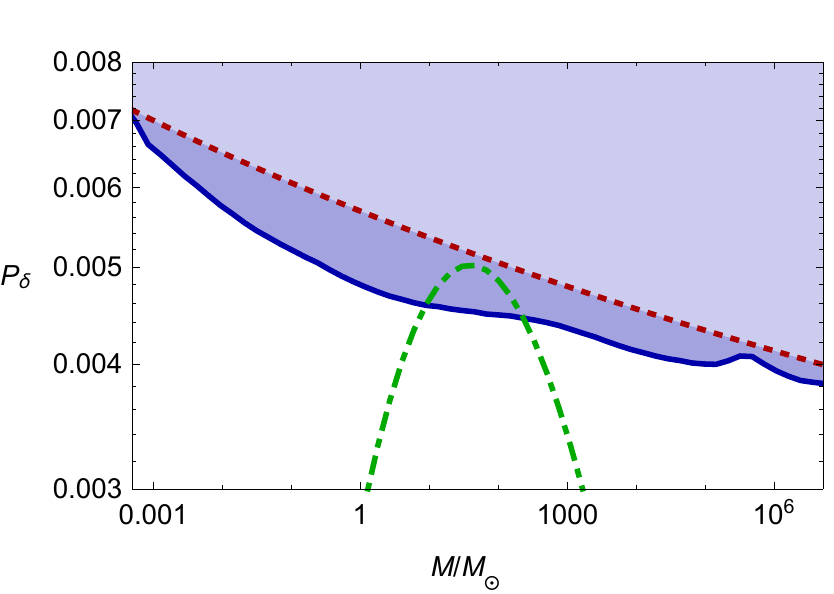}
	\vs{2mm}
	\caption{Would-be power-spectrum constraints (blue, solid curve) from non-observations of PBHs
		obtained from inverting Eq.~\eqref{eq:betaToP}. 
		These constraints are taken from Ref.~\cite{Carr:2016drx}, 
		but for easier comparison with Ref.~\cite{Garcia-Bellido:2016dkw} 
		constraints from the Eridanus II cluster \cite{Brandt:2016aco} and the CMB \cite{Ricotti:2007au} 
		have been excluded. 
		The red, dotted line depicts an inversion according to Eq.~\eqref{eq:betaToP} of the constraints 
		that stem exclusively from the basic requirement $\Omega \leq 1$.
		The green, dot-dashed curve shows the power spectrum of the running-mass model 
		described in the main text.
		}
	\label{fig:PowerSpectrumConstraint}
\end{figure}

In Fig.~\ref{fig:PowerSpectrumConstraint} we show the power-spectrum constraints obtained na{\"i}vely by inverting Eq.~\eqref{eq:betaToP} (excluding the blue-shaded regions above the blue, solid curve). Input constraints on $\beta$ are taken from Ref.~\cite{Carr:2016drx}, but for easier comparison with Ref.~\cite{Garcia-Bellido:2016dkw} constraints from the Eridanus II cluster \cite{Brandt:2016aco} and the CMB \cite{Ricotti:2007au} have been excluded. The green, dot-dashed curve depicts the power spectrum for the running-mass inflationary model specified above (with parameters $\lambda_{1} = 0.011$, $\lambda_{2} = 0.0245$, and $\lambda_{3} = -\.0.00304345$, in order to yield a dark-matter fraction $f$ of $100\%$, peaked around $30\,M_{\odot}$). Seemingly, this model defies the bounds. However, when a PBH mass spectrum is obtained for this model using the critical collapse, it is not excluded by the original PBH constraints. This can be found by applying the critical collapse effect to the power spectrum, as described above. As the critical collapse associated with the PBH formation particularly lowers and broadens any {\it initial} mass spectrum, the {\it resulting} mass distribution will still be compatible with the observational constraints. Here, we have compared the resulting PBH mass distribution to the constraints using the methodology described in Ref.~\cite{Carr:2016drx}.\footnote{We have made sure that the number of bins used to compare the extended mass function to the data is so extensive that the method here does not suffer any of the problems indicated in Ref.~\cite{Green:2016xgy}. Note also that we use a somewhat less constraining set of observations here. However, the result holds that a power spectrum excluded using na{\"i}vely the inverse of Eq.~\eqref{eq:betaToP} on a set of observational constraints on PBHs may still be allowed when critical collapse is applied to obtain the mass distribution. I.e., we do not claim that this PBH-generating power spectrum is generically allowed by all current observations, only by observations we used to originally exclude it at the level of the power spectrum.} Concretely, this allows one to approximately determine, from the constraints calculated assuming a delta-function halo fraction, whether an extended mass function is allowed or not. It utilizes binning of the relevant mass range. Specifically, a given constraint is first divided into locally monotonic pieces. In each of these pieces one starts with the bin, say $i$, where the constraint is smallest, and integrates $\d f / \d M$ over this bin in order to obtain the fraction $f_{i}$ within this bin, to see if it falls below the constraint. Then one goes to the next bin, integrating over $[ M_{i},\.M_{i + 1} ]$, and adds this to $f_i$ in order to obtain $f_{i + 1}$, and so on. By making the bins sufficiently small, any error related to this discrete procedure can be made arbitrarily small.

A potential criticism of this example could be that even though the power spectrum is allowed to have larger amplitudes than the current observational bounds when the critical collapse is taken into account, it cannot go too far from the bounds and therefore the current constraints can still be used with relatively small uncertainties on them, implying that the uncertainties do not affect the observational bounds significantly. It may then be concluded that the non-observation constraints have already excluded a large part of the ``allowed'' region and are therefore useful even if they are not accurate. We however argue that this is not the case, by showing also the constraints computed through the inversion, according to Eq.~\eqref{eq:betaToP}, of the constraints stemming exclusively from the basic requirement $\Omega \leq 1$ for the dark-matter density in the form of PBHs (red, dotted line in Fig.~\ref{fig:PowerSpectrumConstraint}). Even before applying any constraints from the non-observations of PBHs, one can already exclude a large range of the power-spectrum amplitudes at any scales by only applying the $\Omega \leq 1$ condition. This means that we {\it a priori} already have strong constraints on the spectrum, which significantly reduces the region excluded by the non-observation constraints. This remaining region can then be avoided, at least over parts of the mass range, by taking into account the critical-collapse effect, and we therefore do not gain much by adding the non-observation constraints to the original theoretical ones. In addition, the critical-collapse effect smears out the local details of the non-observation constraints, which are commonly used for excluding power spectra with localized features.

In Ref.~\cite{Bugaev:2008gw} a similar analysis has been made, and it is pointed out that the uncertainty due to the critical collapse is considerably smaller than the one coming from the uncertainty in the collapse threshold $\delta_{c}$. This is true to an even larger extent when the possibility of non-spherical collapse is considered. Such an effect will in practice introduce a much larger such uncertainty, though this is a one-way uncertainty only weakening the bounds on the primordial power spectrum. However, we feel that the ramifications of the critical collapse are also important. At present, observational effects are translated into constraints on the power spectrum with localized details recognizable from the observational constraints. As the critical collapse smears out the picture, these details are not necessarily retained in a true rendition of the constraints.

\begin{figure}
	\vs{-3mm}
	\centering
	\includegraphics[scale=1,angle=0]{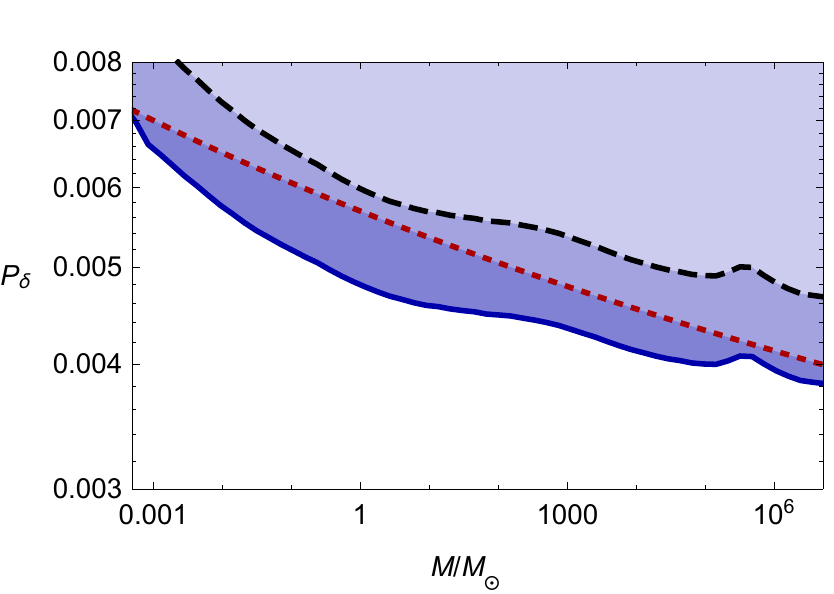}
	\vs{2mm}
	\caption{Would-be power-spectrum constraints (blue, solid curve) from non-observations of PBHs
		obtained from inverting Eq.~\eqref{eq:betaToP}. 
		These are the same as in Fig.~\ref{fig:PowerSpectrumConstraint}.
		The black, dashed curve shows how these constraints will be weakened 
		with a maximal hypothesis of non-sphericity effects in Eq.~\eqref{eq:Non-Spher}.
		Also shown (red, dotted line) is the $\Omega \leq 1$ constraint.
		Note that the latter lies below the former over the entire depicted mass range.
		}
	\label{fig:PowerSpectrumConstraint2}
\end{figure}

We finally turn to non-sphericities, which induce much larger effects on the combined power-spectrum constraints than most of those derived from non-observations of PBHs, such as microlensing constraints. This can be seen clearly in Fig.~\ref{fig:PowerSpectrumConstraint2} in which we again display the same constraints as in Fig.~\ref{fig:PowerSpectrumConstraint} obtained from inverting Eq.~\eqref{eq:betaToP}. Here, we have shown (black, dashed curve) how these constraints will be weakened with a maximal hypothesis of non-sphericity effects in Eq.~\eqref{eq:Non-Spher}. Note that, over the entire depicted mass range, the effect of non-sphericities (which lies much above even the basic constraints stemming exclusively from the basic requirement $\Omega \leq 1$) might be significantly larger than {\it any} of the constraints derived from non-observations of PBHs (microlensing, etc.). Hence, the applicability of these constraints to constrain the primordial power spectrum is invalidated.\footnote{This is even more true when one considers the dependence of the threshold value $\delta_{c}$ on the shape of the overdensity, which can lead to a $50\%$ change of $\delta_{c}$ in some cases (\cf~Refs.~\cite{Shibata:1999zs, Polnarev:2006aa}), making it much larger than the non-spherical effects.} Note that the constraints here coming from the overall bound on $\Omega$ are still obtained using only the simplest na{\"i}ve inversion procedure Eq.~\eqref{eq:betaToP}. We recommend to use this line at present to display a rough level of the constraints today, not for its accuracy, but precisely because its lack of localized features reflects the lack of our current knowledge of the exact processes that lead from the primordial power spectrum to PBHs.

\para{Conclusions}Primordial black holes could form from very high inflationary overdensities. This means that constraints on the existence and abundance of PBHs could yield constraints on the inflationary power spectrum. In principle, these constraints could reach over a huge range of scales inaccessible by CMB and other cosmological observations. However, at present, there is no easy passage from a constraint on the mass spectrum of PBHs. Though such a passage seems possible from na{\"i}ve assumptions, effects like the critical collapse{\,---\,}which is well known{\,---\,}must be taken into account, and at present no route has been made which encompasses these. The critical collapse affects the details of the constraints, making them unreliable. Furthermore, by using the known uncertainties associated with non-spherical effects, we have demonstrated that these might be so large that most of the constraints from non-observations of PBHs would be practically nonexistent. We have shown this by comparing the non-sphericity uncertainties to the improvements in the bounds on the power spectrum that we currently gain by adding non-observation constraints to the constraints coming exclusively from the simple requirement $\Omega \leq 1$ on the dark-matter density in the form of PBHs. A similar comparison shows that the critical-collapse effects can rule in, at least over some scales, power spectra which are currently ruled out, with amplitudes all the way to the $\Omega \leq 1$ bound. In addition, other caveats in the formation and subsequent evolution of PBHs from the inflationary power spectrum, such as the effects of accretion, clustering, or merging of the PBHs after their formation, are currently not fully under control. Until these effects have been studied carefully and taken into account, we recommend to not make detailed constraints on the inflationary power spectrum from non-observation constraints on PBHs. In summary, the current (PBH non-observation) data seem to not be adding much to our ``theoretical priors" on the power spectrum coming from the condition $\Omega \leq 1$, if the uncertainties are taken into account. Therefore, we suggest the cautious scientist to instead apply the simplest scheme for going from PBH constraints to the ones on the primordial power spectrum using only the overall maximum dark-matter constraint. This, although subject to the same uncertainties as non-observation constraints, provides an approximate constraint reflecting our current level of knowledge of mapping the primordial power spectrum to the PBH abundance, without overstating the detailed properties of the constraints. Finally, we believe that our arguments also apply to the constraints placed on the power spectrum by observational constraints on the existence and abundance of UCMHs, as the mechanism behind their formation and evolution resembles that of PBHs. This does not concern the critical collapse, but{\,---\,}even more pronounced{\,---\,}certainly non-sphericities, which may have an even larger effect on the formation of ultra-compact minihalos of dark matter compared to PBHs.

\begin{acknowledgments}
It is a pleasure to thank Christian Byrnes, Jens Chluba, Anne Green, and Peter Klimai for useful comments on a previous version of the manuscript. We also thank Ilia Musco for pointing out to us the large effect of the shape dependence of the overdensities. Y.A. acknowledges support from the Netherlands Organisation for Scientific Research (NWO) and the Dutch Ministry of Education, Culture and Science (OCW), and also from the D-ITP consortium, a program of the NWO that is funded by the OCW. F.K.~acknowledges support from the Vetenskapsr{\aa}det (Swedish Research Council) through contract No.~638-2013-8993 and the Oskar Klein Centre for Cosmoparticle Physics. The work of M.S.~is funded by NORDITA.
\end{acknowledgments}

\bibliography{refs}

\end{document}